# Intrinsic grain boundary mobility tensor from three-dimensional interface random walk

Xinyuan Song and Chuang Deng*

Department of Mechanical Engineering, University of Manitoba, Winnipeg, MB R3T 2N2, Canada

* Corresponding author: Chuang.Deng@umanitoba.ca

**Abstract**

In recent years, studies have demonstrated that the grain boundary (GB) migration is a three-dimensional (3D) process, characterized by a 3D mobility tensor. In this study, we develop a 3D interface random walk theory to extract the GB mobility and shear coupling tensors at equilibrium state based on the random walk of the GB position. Using this approach, we mathematically prove the symmetry of the GB mobility tensor in the case of overdamped GB migration. The theory and its conclusions align with molecular dynamics simulation results and disconnection analysis, and the extracted GB mobility and shear coupling tensors reflect the intrinsic GB properties, unaffected by the large driving forces in atomistic simulations. Additionally, we refined the fast adapted random walk (FAIRWalk) method, enabling efficient extraction of the GB mobility tensor from fewer simulations while maintaining high accuracy. Building on this advancement, we conducted an extensive survey of the mobility, shear coupling, and activation energy for the migration of 388 coincidence-site lattice Ni GBs in the Olmsted database. Several intriguing phenomena were observed, including temperature-induced sudden emergence, disappearance, or inversion of shear coupling; GBs with "zero" normal mobility but high shear mobility; and a non-linear relationship between activation energy and mobility. These findings challenge the traditional understanding of GB migration and warrant further investigation.

# 1 Introduction

Grain boundary (GB) migration is a key process in shaping the microstructure of polycrystalline materials, directly impacting their mechanical [1–10] and thermal properties [11]. Despite more than eight decades of extensive research across experimental [12–19], theoretical [20–22], and computational domains [23–32], predicting GB migration remains a considerable challenge due to the intrinsic complexities of GB behavior.

The classic kinetic equation for GB migration is expressed as:

$$v = MP|_{P \to 0} \tag{1}$$

where $v$ is GB velocity, $P$ is external driving force, and $M$ represents the intrinsic GB mobility. Traditional large-scale studies on GB network evolution have typically assumed that all GBs have identical mobility and that their velocities are solely determined by the external driving force. However, recent studies [19,33–36] have increasingly highlighted the diverse nature of GB mobility, challenging this assumption. Recent advancements in computational techniques [37–42] have enabled researchers to investigate GB migration at the atomic scale, and numerous intriguing phenomena related to GB mobility have been revealed over the past decade [43–49]. To capture these phenomena within short simulation times, many studies have relied on applying large driving forces. However, using unrealistically large driving forces can alter the potential energy landscape of GB migration [44] and even drive the GB far from equilibrium [40]. Research has demonstrated a strong dependence of GB mobility on the driving force [44,50,51], as well as driving force-induced transitions in its thermal behavior [44].

Equilibrium methods have been developed to eliminate the effects of driving forces, such as the interface random walk method [40,52] and approaches based on the autocorrelation decay of GB fluctuations [41,42]. The mobilities obtained through these equilibrium methods show excellent agreement with those derived from Eq. 1 in the zero-driving-force limit [40,44,53]. However, these methods rely on extensive statistical analysis of GB fluctuations over time, requiring a large volume of data, which is computationally intensive. This is one of the reasons why equilibrium methods have not been widely adopted in large-scale surveys of GB mobility [33,34,43,54].

Furthermore, current equilibrium methods [40–42,52] are predominantly based on 1D models. For example, as illustrated in Fig. 1a, only the component of the bulk Langevin force $\eta$ associated with the direction of GB fluctuation is considered. In reality, $\eta$ should be treated as a 3D vector, and due to the shear coupling effect of GBs—where adjacent grains undergo synchronized displacements alongside the normal migration of the GB (referred to as GB shear displacement)—the GB migration can also be driven by forces in other directions, as shown in Fig. 1b.

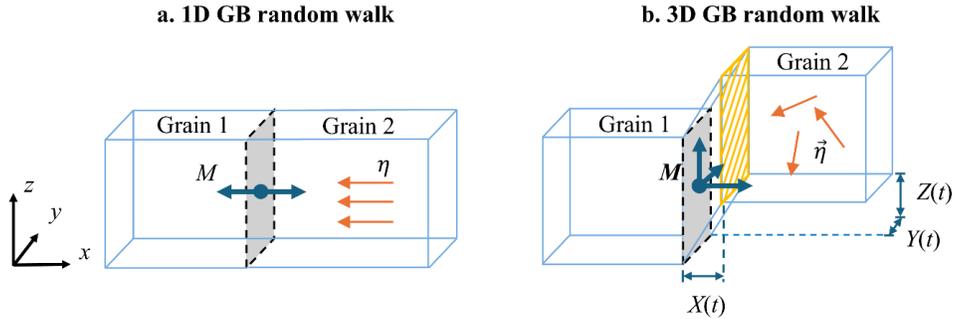

Figure 1 Diagram of (a) 1D interface random walk model, and (b) 3D interface random walk model.

Recently, Chen et al. [55] proposed the concept of GB mobility tensor, expressed as

$$\begin{bmatrix} v_x \\ v_y \\ v_z \end{bmatrix} = \begin{bmatrix} M_{xx} & M_{xy} & M_{xz} \\ M_{yx} & M_{yy} & M_{yz} \\ M_{zx} & M_{zy} & M_{zz} \end{bmatrix} \begin{bmatrix} \varphi \\ \tau_y \\ \tau_z \end{bmatrix} \qquad (2)$$

where, $\varphi$ is the driving force normal to the GB plane, and $\tau_y$ and $\tau_z$ are shear driving forces tangential to the GB plane. The GB mobility tensor provides a more comprehensive representation of 3D GB kinetics, making it especially useful for studying complex GB migration within intricate GB networks [56,57]. In this study, our goal is to develop a 3D interface random walk (3D-IRW) theory to extract the intrinsic GB mobility tensor from the equilibrium state of the GB. By integrating this approach with an efficient data processing technique [53], we conduct an extensive survey of GB mobility tensors across 388 Olmsted coincidence-site lattice (CSL) Ni GBs [58].

Shear coupling is another critical phenomenon observed during GB migration. It has traditionally been characterized by the shear coupling factor, defined as the ratio of GB displacement or velocity in the shear direction to that in the normal direction, i.e., $\beta = d_z/d_x$ or $v_z/v_x$. However, $\beta$ has been shown to exhibit a strong dependency on the driving force [30], suggesting that previous large driving forces-based survey may have produced results that are artificial and deviated from reality. A recent concept, the intrinsic shear coupling tensor [53], extracted from the GB mobility tensor, has demonstrated excellent predictive ability regarding the variation of the apparent $\beta$ with driving force. In this study, we further explore the theoretical background of the shear coupling tensor using the newly developed 3D-IRW theory. Additionally, we conduct an extensive survey of the shear coupling tensor across the 388 Olmsted CSL Ni GBs [58].

## 2 Continuum theories

The equilibrium equation for a GB under thermal fluctuations [59–61] can be extended to 3D scenarios as follows :

$$\langle \vec{\eta}(\vec{r},t)\vec{\eta}(\vec{r}',t')^T \rangle = 2k_B T \boldsymbol{\Gamma} \delta(\vec{r}-\vec{r}')(t-t') \tag{3}$$

where $\vec{\eta}$ is thermal noise, $\boldsymbol{\Gamma}$ is the dissipation tensor, $\delta$ is Dirac function which applies to both time ($t$) and position, $\vec{r}$ is the coordinate on GB surface. According to the fluctuation-dissipation theorem[40,41,62,63], the GB mobility tensor is the inverse of the dissipation tensor, expressed as:

$$\boldsymbol{M} = \boldsymbol{\Gamma}^{-1} \tag{4}$$

Assuming the GB is planar, the driving force arising from curvature can be neglected. The Brownian motion of a point on the GB is described by the Langevin equation [59,60]:

$$m\frac{d\vec{v}}{dt} = -\boldsymbol{\Gamma}\vec{v} + \vec{\eta}(\vec{r},t) \tag{5}$$

Here, $m$ is the effective GB mass. The Langevin equation describes the motion of an interface as influenced by its inertia ($m\frac{d\vec{v}}{dt}$), a dissipative force ($-\boldsymbol{\Gamma}\vec{v}$) from the surrounding medium, which is proportional to the particle's velocity and acting in the opposite direction, and random thermal fluctuations ($\vec{\eta}(\vec{r},t)$).

### 2.1 Overdamped 3D-IRW theory

The original interface random walk theory assumes that the motion of the GB is overdamped [40]. This implies that each element in the dissipation tensor is much greater than the effective mass, making the relaxation term $\boldsymbol{\tau} = m\boldsymbol{\Gamma}^{-1}$ negligible. As a result, the velocity of the interface can respond almost instantaneously to external fluctuations. Under these conditions, Eq. 5 simplifies to:

$$\vec{v} = \boldsymbol{\Gamma}^{-1}\vec{\eta}(\vec{r},t) = \boldsymbol{M}\vec{\eta}(\vec{r},t) \tag{6}$$

The average velocity, $\vec{\bar{v}}$, of the entire GB can be obtained by integrating both sides of Eq. 6 over the GB surface area $A$:

$$\vec{\bar{v}} = \frac{\boldsymbol{M}}{A} \iint_A \vec{\eta}(\boldsymbol{r}, t) \mathrm{d}\boldsymbol{r} \tag{7}$$

Further integration of Eq. 7 over time leads to the kinetic equation for the GB instant position:

$$\vec{R}(t) = \boldsymbol{M} \int_0^t \iint_A \vec{\eta}(\boldsymbol{r}, t) \mathrm{d}\boldsymbol{r}\, dt' \tag{8}$$

where, $\vec{R}(t) = \begin{bmatrix} X_t \\ Y_t \\ Z_t \end{bmatrix}$, and $X_t, Y_t, Z_t$ are instant GB position at time $t$. By calculating the covariance matrix for both sides in Eq. 8, utilizing Eq. 3, the following relation is ultimately derived:

$$\boldsymbol{\Sigma}_{\vec{R}_t} = \frac{2k_B T t}{A} \boldsymbol{M}^T \tag{9}$$

where

$$\boldsymbol{\Sigma}_{\vec{R}_t} = \begin{bmatrix} Var(X_t) & Cov(X_t, Y_t) & Cov(X_t, Z_t) \\ Cov(X_t, Y_t) & Var(Y_t) & Cov(Y_t, Z_t) \\ Cov(X_t, Z_t) & Cov(Y_t, Z_t) & Var(Z_t) \end{bmatrix} \tag{10}$$

and $\boldsymbol{M}^T$ is the transpose of the GB mobility tensor $\begin{bmatrix} M_{xx} & M_{xy} & M_{xz} \\ M_{yx} & M_{yy} & M_{yz} \\ M_{zx} & M_{zy} & M_{zz} \end{bmatrix}^T$. Eq. 9 unveils a very important principle that, similar to $M_{xx} \propto Var(X) = Cov(X,X)$ and $M_{zz} \propto Var(Z) = Cov(Z,Z)$, the off-diagonal elements in the mobility tensor, exemplified by the x-z plane, $M_{xz} = M_{zx} \propto Cov(X,Z)$ with the same proportional factor $2k_B T t/A$. In this way all the elements in GB mobility tensor can be unified by the same expression:

$$M_{\alpha\gamma} = \frac{A}{2k_B T t} Cov(\alpha, \gamma) \tag{11}$$

where, $\alpha$ and $\gamma$ denote GB migration directions. Eq. 11 demonstrates the structural integrity of the GB mobility tensor. Furthermore, the symmetry of the GB mobility tensor as predicted by Onsager relation [55,64] is now mathematically proved.

## 2.2 General 3D-IRW theory

When the assumption that $\Gamma_{ij} \gg m$ is not satisfied, the time dependence of the response function must be considered. Returning to Eq. 5, the $\boldsymbol{\Gamma}$ can be decomposed into the form $\boldsymbol{P\Lambda P^{-1}}$, where $\boldsymbol{\Lambda} = \begin{pmatrix} \lambda_1 & & \\ & \lambda_2 & \\ & & \lambda_3 \end{pmatrix}$ is a diagonal matrix containing the eigenvalues ($\lambda_i$) of $\boldsymbol{\Gamma}$, and $\boldsymbol{P}$ is the matrix whose columns are the eigenvectors of $\boldsymbol{\Gamma}$. Defining $\vec{u} = \boldsymbol{P}^{-1}\vec{v}$ results in the following expression:

$$m \frac{d\vec{u}}{dt} = -\boldsymbol{\Lambda}\vec{u} + \boldsymbol{P}^{-1}\vec{\eta}(\vec{r}, t) \tag{12}$$

Eq. 12 can be further decomposed into three one-dimensional problems:

$$m \begin{pmatrix} \dfrac{du_1}{dt} \\ \dfrac{du_2}{dt} \\ \dfrac{du_3}{dt} \end{pmatrix} = \begin{pmatrix} \lambda_1 u_1 + \left[\boldsymbol{P}^{-1} \dfrac{\vec{\eta}(\vec{r},t)}{m}\right]_1 \\ \lambda_2 u_2 + \left[\boldsymbol{P}^{-1} \dfrac{\vec{\eta}(\vec{r},t)}{m}\right]_2 \\ \lambda_3 u_3 + \left[\boldsymbol{P}^{-1} \dfrac{\vec{\eta}(\vec{r},t)}{m}\right]_3 \end{pmatrix} \tag{13}$$

For each dimension, we solve the equation using the procedure outlined in Karma et al.'s work[41]. Integrating both sides over time yields:

$$\begin{pmatrix} u_1 \\ u_2 \\ u_3 \end{pmatrix} = \begin{pmatrix} u_{0_1} e^{-\frac{\lambda_1 t}{m}} + \int_0^t e^{-\frac{\lambda_1(t-s)}{m}} \left[\boldsymbol{P}^{-1} \dfrac{\vec{\eta}(\vec{r},s)}{m}\right]_1 ds \\ u_{0_2} e^{-\frac{\lambda_2 t}{m}} + \int_0^t e^{-\frac{\lambda_2(t-s)}{m}} \left[\boldsymbol{P}^{-1} \dfrac{\vec{\eta}(\vec{r},s)}{m}\right]_2 ds \\ u_{0_3} e^{-\frac{\lambda_3 t}{m}} + \int_0^t e^{-\frac{\lambda_3(t-s)}{m}} \left[\boldsymbol{P}^{-1} \dfrac{\vec{\eta}(\vec{r},s)}{m}\right]_3 ds \end{pmatrix} \tag{14}$$

Or in vector form:

$$\vec{u} = e^{-\frac{\boldsymbol{\Lambda} t}{m}} \vec{u}_0 + \int_0^t e^{-\frac{\boldsymbol{\Lambda}(t-s)}{m}} \boldsymbol{P}^{-1} \frac{\vec{\eta}(\vec{r},s)}{m} ds \tag{15}$$

By substituting $\boldsymbol{\Gamma} = \boldsymbol{P\Lambda P^{-1}}$ and $\vec{u} = \boldsymbol{P}^{-1}\vec{v}$ into Eq. 15, the 3D velocity equation for each point on the GB surface can be derived as:

$$\vec{v} = e^{-\Gamma m^{-1} t} \vec{v}_0 + \int_0^t e^{-\frac{\Gamma(t-s)}{m}} \frac{\vec{\eta}(\vec{r}, s)}{m} ds \quad (16)$$

Integrating Eq. 16 over time gives the displacement equation for each point on the GB surface:

$$\vec{x} = \vec{x}_0 + \int_0^t e^{-\frac{\Gamma(t-s)}{m}} ds\, \vec{v}_0 + \int_0^t ds_1 \int_0^{s_1} e^{-\frac{\Gamma(s_1-s_2)}{m}} \frac{\vec{\eta}(\vec{r}, s_2)}{m} ds_2 \quad (17)$$

Further integrating both sides of Eq. 17 over the GB surface area $A$ yields the 3D kinetic equation for the average GB position:

$$\vec{R}_t = \vec{R}_0 + \int_0^t e^{-\frac{\Gamma(t-s)}{m}} ds\, \vec{\bar{v}}_0 + \frac{1}{A} \iint_A d\vec{r} \int_0^t ds_1 \int_0^{s_1} e^{-\frac{\Gamma(s_1-s_2)}{m}} \frac{\vec{\eta}(\vec{r}, s_2)}{m} ds_2 \quad (18)$$

By calculating the covariance matrix for both sides of Eq. 18, one can disregard the constant vectors, as the covariance of any vector with a constant vector is zero. This leads us to derive the following relation:

$$\begin{aligned}
\Sigma_{\vec{R}_t} &= \langle \frac{1}{A} \iint_A d\vec{r} \int_0^t ds_1 \int_0^{s_1} e^{-\frac{\Gamma(s_1-s_2)}{m}} \frac{\vec{\eta}(\vec{r}, s_2)}{m} ds_2, \left[\frac{1}{A} \iint_A d\vec{r}' \int_0^t ds_1' \int_0^{s_1'} e^{-\frac{\Gamma(s_1'-s_2')}{m}} \frac{\vec{\eta}(\vec{r}, s_2)}{m} ds_2'\right]^T \rangle \\
&= \frac{1}{A^2 m^2} \iint_A d\vec{r} \iint_A d\vec{r}' \int_0^t ds_1 \int_0^t ds_1' \int_0^{s_1} \int_0^{s_1'} e^{-\frac{\Gamma(s_1-s_2)}{m}} \langle \vec{\eta}(\vec{r}, s_2) \vec{\eta}(\vec{r}', s_2')^T \rangle e^{-\frac{\Gamma^T(s_1'-s_2')}{m}} ds_2\, ds_2'
\end{aligned} \quad (19)$$

Introducing Eq. 3 into Eq. 19, one can get

$$\begin{aligned}
\Sigma_{\vec{R}_t} &= \frac{2 k_B T}{A^2 m^2} \iint_A d\vec{r} \iint_A d\vec{r}' \int_0^t ds_1 \int_0^t ds_1' \int_0^{s_1} \int_0^{s_1'} e^{-\frac{\Gamma(s_1-s_2)}{m}} \Gamma \delta(\vec{r}-\vec{r}')(s_2-s_2') e^{-\frac{\Gamma^T(s_1'-s_2')}{m}} ds_2\, ds_2' \\
&= \frac{2 k_B T}{A m^2} \int_0^t ds_1 \int_0^t ds_1' \int_0^{s_1} \int_0^{s_1'} P e^{-\frac{\Lambda(s_1-s_2)}{m}} P^{-1} P \Lambda P^{-1} (P^{-1})^T e^{-\frac{\Lambda(s_1'-s_2')}{m}} P^T \delta(s_2-s_2') ds_2 ds_2'
\end{aligned} \quad (20)$$

To simplify the equation, it is assumed that $\Gamma$ is symmetric, which implies $P^{-1} = P^T$, and $P^{-1} P = P^{-1}(P^{-1})^T = I$. This assumption simplifies Eq. 20 into three 1D random walk cases:

$$\Sigma_{\vec{R}_t} = \frac{1}{A} P \begin{pmatrix} \frac{2 k_B T}{m^2} \int_0^t ds_1 \int_0^t ds_1' \int_0^{s_1} \int_0^{s_1'} e^{-\frac{\lambda_1(s_1-s_2)}{m}} \lambda_1 e^{-\frac{\lambda_1(s_1'-s_2)}{m}} \delta(s_2-s_2') ds_2 & & \\ & V(\lambda_2) & \\ & & V(\lambda_3) \end{pmatrix} P^T \quad (21)$$

$$= \frac{1}{A} \boldsymbol{P} \begin{pmatrix} \frac{2k_BT}{\lambda_1}\left(t - \tau\left(1 - e^{-\frac{t}{\tau}}\right)\right) & & \\ & V(\lambda_2) & \\ & & V(\lambda_3) \end{pmatrix} \boldsymbol{P}^T$$

where $\tau = m/\lambda_i$ is the relaxation time and $V(\lambda_i)$ represents a similar expression to the first element of the tensor, adjusted for different $\lambda_i$. In the long-time limit ($t \gg \tau$), each 1D random walk equation can be simplified to $V(\lambda_i) = \frac{2k_BT}{\lambda_i}t$, which is the Einstein equation [41,61,63,65–67] for the Brownian motion of a free particle. Therefore, Eq. 21 can be eventually simplified to

$$\boldsymbol{\Sigma}_{\vec{R}_t} = \frac{2k_BTt}{A}(\boldsymbol{\Gamma}^{-1})^T = \frac{2k_BTt}{A}\boldsymbol{M}^T \qquad (22)$$

Eq. 22 yields the same result as the overdamped 3D-IRW (Eq. 9). The derivation is based on the crucial assumption that the GB dissipation or mobility tensor is symmetric. This symmetry, as predicted by the Onsager relation [55,64], has been confirmed by the MD simulations in the previous study [53].

The interface random walk theory relies on the key assumption that the GB is planar. However, the study by Karma et al. [41] demonstrates that the interface random walk method remains applicable even for large and fluctuating GBs.

## 3. MD simulations

The models used in this study are from the Olmsted 388 Ni CSL GB database [58]. A free surface boundary condition is applied normal to the GB plane ($x$ direction), while periodic boundary conditions are used in the directions parallel to the GB plane ($y$ and $z$ directions). A typical model and its dimensions are depicted in Fig. 2. Before the simulations, all the models are first expanded based on the thermal expansion coefficients at different temperatures and then equilibrated under zero pressure with an isothermal–isobaric (NPT) ensemble for 300 ps, followed by 100 ps of relaxation using a canonical ensemble (NVT) with a Berendsen thermostat [68]. Afterward, the thermostat is removed, allowing the GBs to undergo a random walk under a microcanonical (NVE) ensemble for 600 ps. Each simulation is repeated 50 times at each temperature with different initial velocity distributions. To maximize simulation time while minimizing computational cost, a larger timestep is preferred. However, at low temperatures, large timesteps may fail to capture subtle atomic movements, leading to significant truncation errors in the Verlet algorithm [69,70].

Following the recommendation in Ref. [70], we used two different timesteps: 1 fs for the low-temperature regime (10–200 K) and 5 fs for the room-to-high-temperature regime (300–1500 K). A detailed explanation of the timestep selection and its validity is provided in Section S1 of the Supplemental Materials.

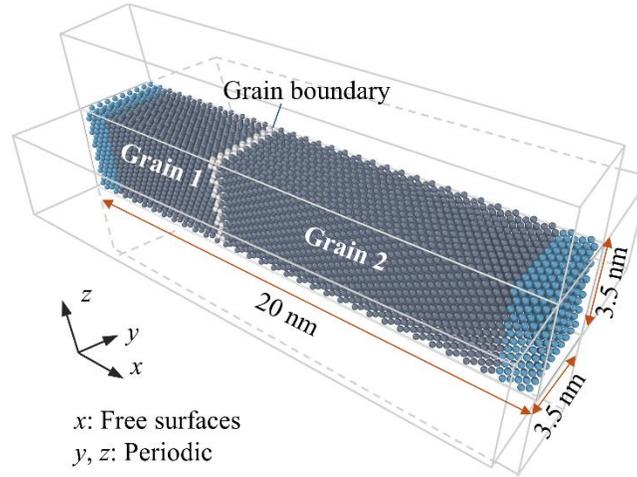

Figure 2 A typical model in Olmsted 388 Ni GB database [58]

During the simulation, GB normal displacement is tracked using the order parameter defined in Ref. [38]. To monitor GB shear displacement, the common neighbor analysis (CNA) method [71,72] is employed to track the instantaneous shear position of the GB. Additionally, the relative displacement between two slabs at opposite ends of the model (blue areas in Fig. 2) is recorded. These two methods differ in the masses ($m$) considered in Eq. 5: for the instant GB shear displacement (tracked using CNA), only the effective mass of the GB is considered, whereas for the displacement of the slabs, the mass of the two grains is also included. According to fluctuation–dissipation theorem [61–63], the relaxation time for a freely moving object responding to a fluctuating force is proportional to the mass involved, i.e., $\tau = m/\mu$, where $\mu$ is the dissipation. This suggests there should be a delay in the displacement of the slabs compared to the instant GB shear displacement. However, as shown in Fig. 3, the shear displacements tracked by both methods are synchronized, supporting the assumption that GB migration is overdamped ($\mu \gg m$), allowing the mass term to be neglected. Since the CNA method requires continuous output of configuration files to track GB migration, we used the relative displacement of the two slabs as a proxy for GB shear displacement in large-scale surveys.

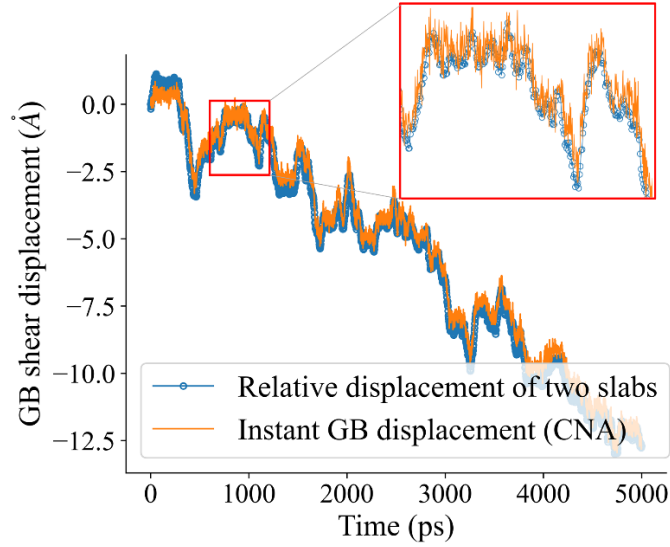

Figure 3 Synchronized displacements of the instant GB shear position and the two slabs at two ends of the model.

In the force-driven GB migration simulations, the normal driving force ($\varphi$) is applied by imposing a synthetic energy jump across the GB [37–39], while the shear driving forces ($\tau$) are introduced by applying opposite forces to two slabs positioned at the ends of the model (blue areas in Fig. 2). Each simulation is repeated 20 times with different initial velocity distributions. The GB velocity is then calculated by averaging the results from these 20 simulations, and the GB mobility tensor is determined using Eq. 2.

All simulations were conducted using the Large-scale Atomic/Molecular Massively Parallel Simulator (LAMMPS) [73]. The interatomic interactions are calculated using an embedded atom method (EAM) potential [74] specifically designed for Ni [75]. The GB configurations are visualized through the Open Visualization Tool (OVITO) [76].

**3.1 Fast adapted interface random walk (FAIRWalk) and intrinsic GB shear coupling tensor**

According to Eq. 9 or 22, the principal mobilities in the GB mobility tensor can be determined using the equation [40,41]:

$$\sigma_t^2 = Dt \tag{23}$$

where, $\sigma_t^2$ is the instant variance of GB displacement at moment $t$, and $D$ is a diffusional coefficient related to GB migration. The GB mobility can then be calculated as

$$M = \frac{DA}{2k_B T} \qquad (24)$$

In our previous study [53], we established a linear relationship between the variance of cumulative GB displacement, $\sigma_c^2$, and time $t$:

$$\sigma_c^2 = \frac{1}{2} Dt \qquad (25)$$

The advantage of using $\sigma_c^2$ is that it incorporates all available data simultaneously, rather than relying on instantaneous data at specific moments to determine GB mobility. This significantly enhances statistical accuracy without requiring additional simulations. As shown in Fig. 4a, $\sigma_c^2$ exhibits a much stronger linear relationship with $t$ compared to $\sigma_t^2$. Fig. 4b further demonstrates that, with the same number of parallel simulations (50 simulations), the GB mobility calculated using $\sigma_c^2$ converges to a stable value at around 100 ps, which is considerably faster than the conventional interface random walk method. This result also suggests that our choice of simulation parameters—50 independent simulations, each running for 600 ps—should be sufficient to obtain reliable data.

A similar data processing idea is employed in the adapted interface random walk (AIRWalk) method proposed by Deng and Schuh [52], where each simulation is divided into multiple segments. To improve statistical accuracy, the data is reused in each segment, as shown in Fig. 4c. In the extreme case where each segment contains nearly all the data, the calculated mobility would approach to the value obtained from $\sigma_c^2$. However, the computation time for AIRWalk is proportional to the number of segments, often taking hours to process the data in practice. In contrast, the $\sigma_c^2 - t$ curve can be generated in just a few minutes, which is why we refer to this method as the fast adapted interface random walk (FAIRWalk) method.

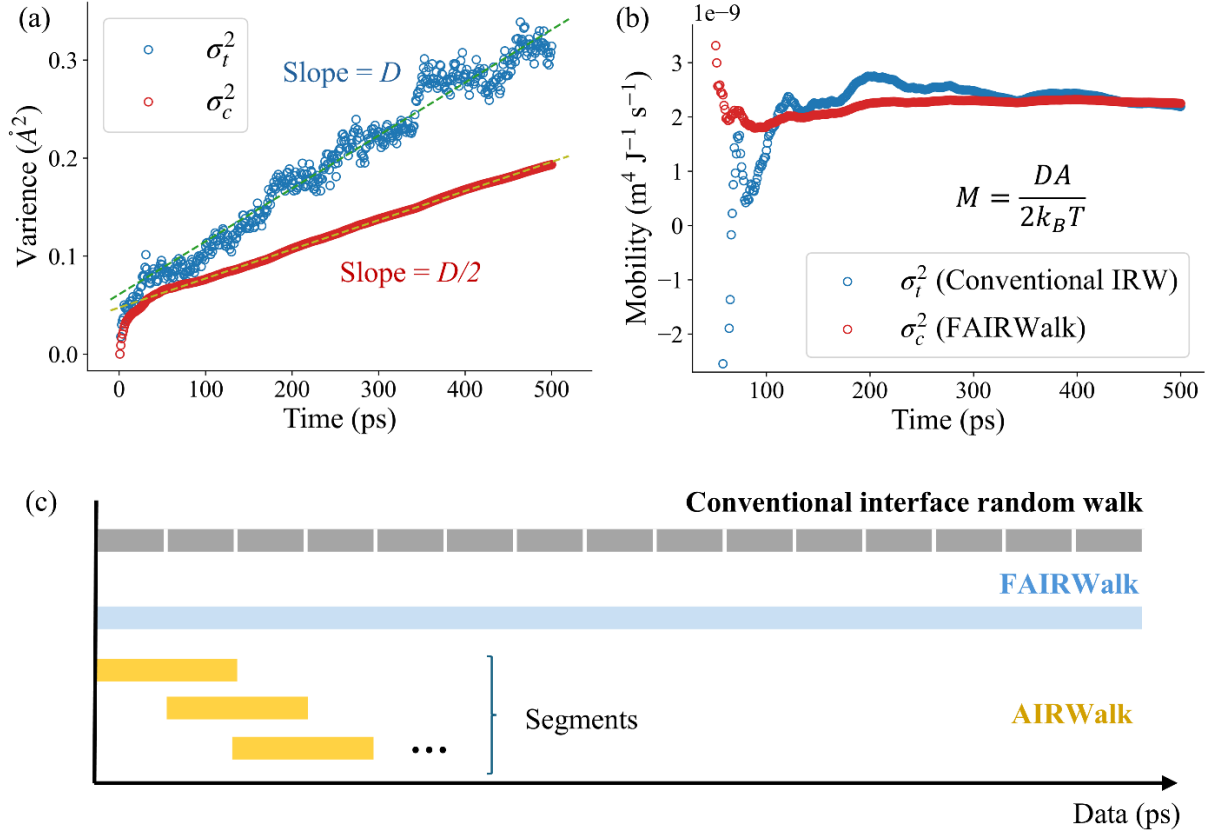

Figure 4 Plots of (a) $\sigma_t^2$ and $\sigma_c^2$ vs. $t$ curves and (b) $M$ vs. $t$ obtained using different data processing method. All the data are obtained from interface random walk simulations of Σ5 (3 1 0) Ni GB (p1 in Olmsted database [58]) at 1000K. Subfigure (a) also highlights that the initial steep region should be excluded from the analysis, regardless of the method used. (c) Schematic of the data processing methods used in conventional interface random walk [40], AIRWalk [52], and FAIRWalk [53].

We now aim to extend the FAIRWalk method to quickly calculate the off-diagonal elements in GB mobility tensor. By utilizing the accumulated GB displacements in the 3D directions, i.e. $X_c$, $Y_c$, and $Z_c$, Eqs. 9 and 22 are revised to:

$$\boldsymbol{\Sigma}_{\vec{R}_c} = \frac{k_B T t}{A} \boldsymbol{M}^T \tag{26}$$

where $\boldsymbol{\Sigma}_{\vec{R}_c}$ is the covariance tensor of the accumulated GB displacements across different directions. By closely examining Eq. 26, we find that the ratio of off-diagonal elements in the GB mobility tensor to the principal mobilities can be expressed as follows (using $M_{zx}$ and $M_{xx}$ as an example):

$$\frac{M_{zx}}{M_{xx}} = \frac{\text{Cov}(X_c, Z_c)}{\text{Var}(X_c)} = \rho \frac{\sigma_z}{\sigma_x} = S_{zx} \tag{27}$$

Here, $\rho$ is Pearson correlation coefficient, $\sigma_z^2$ and $\sigma_x^2$ represent cumulated GB displacements in $z$ and $x$ directions, respectively. The combination of these terms forms the expression for the fitted slope ($S_{zx}$) obtained through simple linear regression, which we previously defined as the shear coupling strength in our study on GB shear coupling [53]. This means the relationship between elements of the GB mobility tensor can be determined by plotting cumulative GB displacements on a single graph and performing simple linear regression to fit $S$. Examples of fully coupled and partially coupled GB migration, along with their fitted $S$, are shown in Figs. 5a and b, respectively.

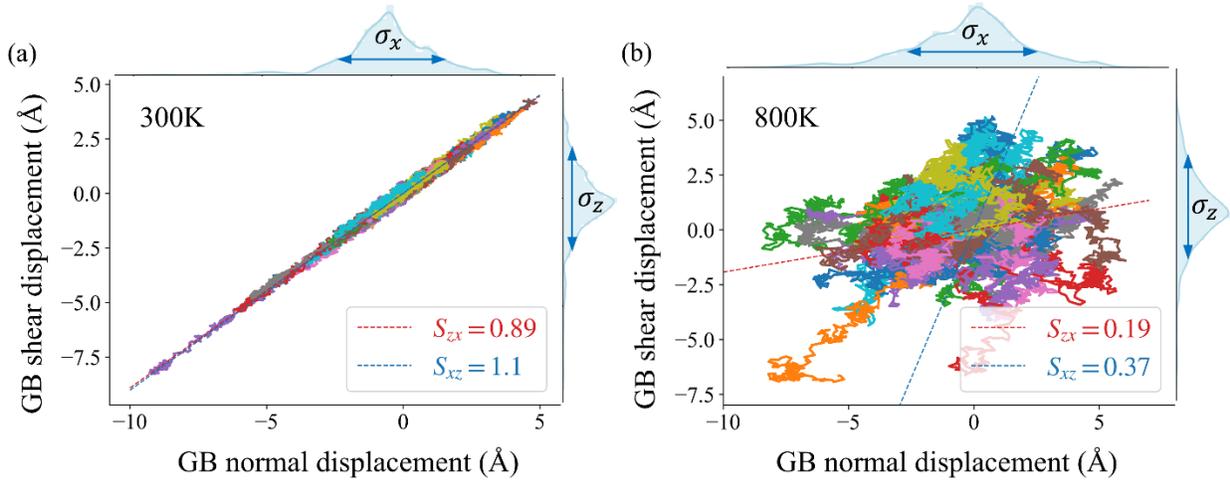

Figure 5 Plots of cumulated GB displacements in $x$ and $z$ directions for $\Sigma 15$ (2 1 1) GBs (p14 in Olmsted database [58]), at 300K (fully coupled) and 800K (partially coupled). $S_{zx}$ and $S_{xz}$ are obtained by applying simple linear regression to the $X$-$Z$ and $Z$-$X$ curves, respectively. Different colors represent data from different independent simulations. At low temperature (300K), the GB normal and shear displacements exhibit a strong linear correlation, indicating strong shear-coupled GB migration in both directions. In contrast, at high temperature (800K), this correlation weakens, and the GB migration becomes decoupled.

Therefore, the GB mobility tensor can be also expressed as

$$\begin{bmatrix} M_{xx} & M_{xy} & M_{xz} \\ M_{yx} & M_{yy} & M_{yz} \\ M_{zx} & M_{zy} & M_{zz} \end{bmatrix} = \begin{bmatrix} 1 & S_{xy} & S_{xz} \\ S_{yx} & 1 & S_{yz} \\ S_{zx} & S_{zy} & 1 \end{bmatrix} \begin{bmatrix} M_{xx} & & \\ & M_{yy} & \\ & & M_{zz} \end{bmatrix} \tag{28}$$

Here, the $S$ tensor is referred to as the intrinsic shear coupling tensor [53].

## 3.2 Verification of the 3D-IRW theory

Numerous studies [40,41,44,53] have confirmed the ability of the interface random walk method to capture the diagonal elements of the GB mobility tensor, i.e., $M_{xx}$, $M_{yy}$, and $M_{zz}$, in the zero-force limit. A key advantage of the 3D-IRW theory is its ability to extract off-diagonal elements from the same simulations. For example, $M_{zx}$ represents the GB mobility in the $z$-direction when subjected to a driving force in the $x$-direction. Traditionally, obtaining this value requires applying an external force. However, according to Eq. 26 (or Eqs. 9 and 22), this shear-coupled mobility can be derived by calculating the covariance of displacements, Cov($X,Z$), from interface random walk simulations at zero-driving force limit. This approach allows all elements of the GB mobility tensor to be obtained from the same set of simulation data, enabling further exploration of the intrinsic relationships between elements without the influence of different types of driving forces.

To validate the 3D-IRW theory, the FAIRWalk method (Eq. 26) was applied to Σ13 (5 1 0), Σ37 (7 5 0), and Σ15 (2 1 1) GBs (p7, p61 and p14 in Olmsted database [58]), which exhibit shear coupling in the $x$-$z$ plane and have been extensively studied in previous literature [30,53]. Additionally, Σ15 (5 2 1) GB (p9 in Olmsted database [58]), which shows evident shear coupling in all three directions, was included. Fig. 6 demonstrates that Eq. 26 effectively captures shear-coupled mobilities. The results also confirm the symmetry of the GB mobility tensor, as predicted by the overdamped 3D-IRW theory (Eq. 9). It is worth noting that the error bars in Fig. 6 are obtained from 20 independent force-driven GB migration simulations. The large error bars at high temperatures are primarily due to significant thermal fluctuations and weakened shear coupling. The off-diagonal elements in the GB mobility tensor represent shear-coupled mobilities. For instance, $M_{xz}$ corresponds to GB normal mobility under shear stress. At high temperatures, the shear coupling weakens, as evidenced by the very low correlation coefficients shown in Fig. 5. Consequently, the effect of shear stress (driving force) becomes minimal, and under significant thermal fluctuations, GB migration resembles random walk behavior. To extract reliable GB mobility under these conditions, extensive simulations are required to achieve high statistical accuracy. Moreover, obtaining other elements of the GB mobility tensor necessitates applying different types of driving forces and conducting additional groups of simulations.

In this aspect, the FAIRWalk method is significantly more efficient than traditional force-driven approaches. It not only captures weakened shear coupling information effectively but also enables the extraction of all elements of the GB mobility tensor simultaneously from a single group of simulations.

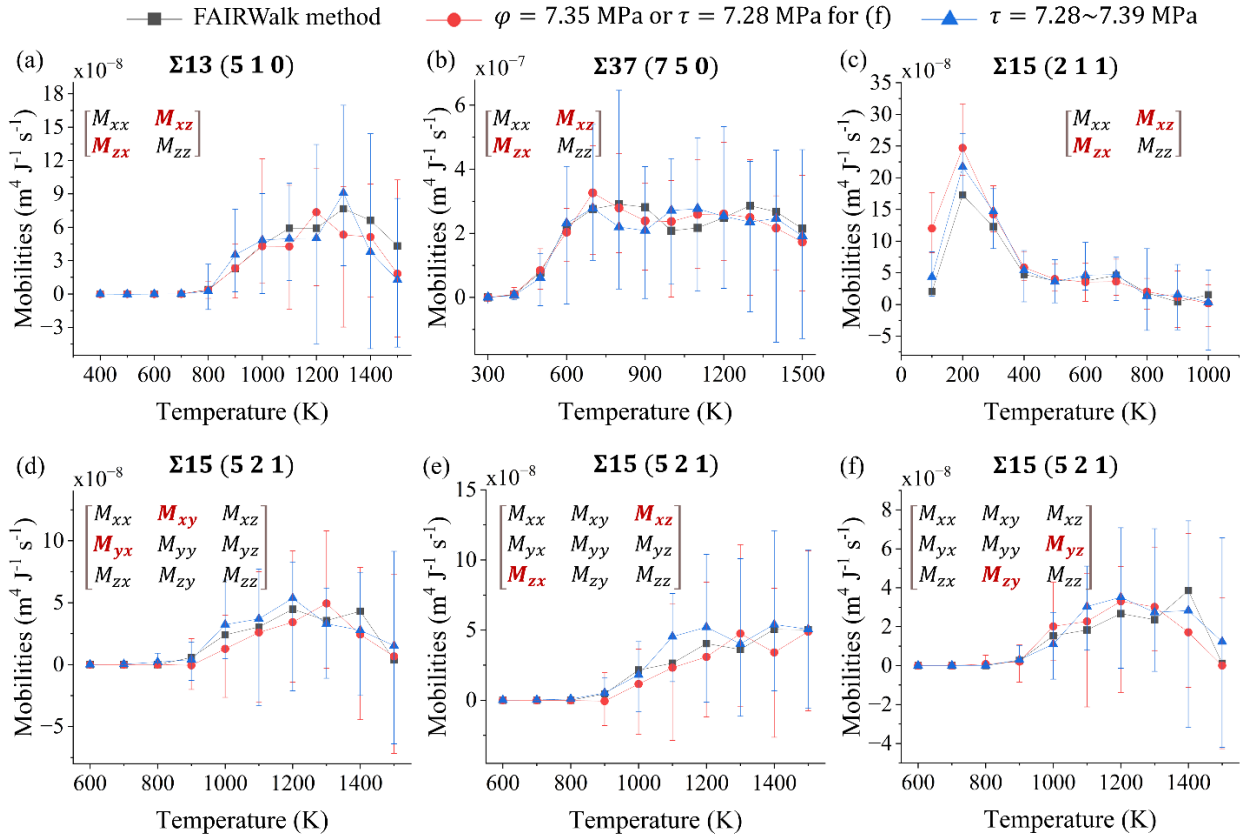

Figure 6 Shear-coupled mobilities (off-diagonal elements of the GB mobility tensor) determined using the FAIRWalk method (Eq. 27) for Σ13 (5 1 0), Σ37 (7 5 0), Σ15 (2 1 1), and Σ15 (5 2 1) GBs. The results are compared to mobilities calculated from simulations with external driving forces (Eq. 2). The large error bars at high temperatures are primarily due to significant thermal fluctuations and weakened shear coupling.

Additionally, the product of $S$ at the symmetric position in the shear coupling tensor (e.g., $S_{zx}$ and $S_{xz}$), which equals the square of the Pearson coefficient ($\rho^2$), ranging from 0 to 1, can serve as an indicator of the coupling state of GB migration. We refer to $\rho^2$ as the standard coupling factor. As shown in Fig. 7, when $\rho^2_{zx} = 1$, GB migration is fully coupled, meaning the displacement in $x$ or $z$ direction will always result in a fixed displacement in the other direction, as illustrated in Fig. 5a. This typically occurs in low-temperature regimes. Conversely, when $\rho^2_{zx} = 0$, the GB is uncoupled, indicating that displacements in $x$ and $z$ directions are entirely independent. Since there is no shear coupling in the *x-y* and *y-z* planes of the Σ15 (2 1 1) Ni GB, $\rho_{yx}^2$ and $\rho_{zy}^2$ remain near zero across all temperatures.

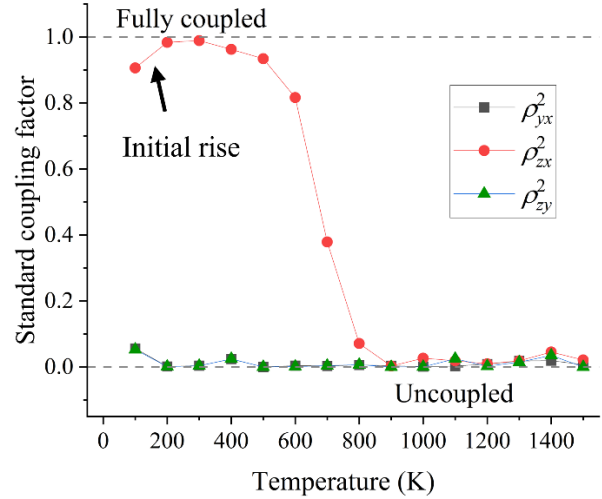

Figure 7 Variation of the standard coupling factor $\rho^2$ with temperature for the Σ15 (2 1 1) Ni GB.

The initial rise in $\rho_{zx}^2$ from 100K to 200K, as shown in Fig. 7, is a common phenomenon observed in our subsequent survey of 388 GBs. This rise is not due to the GB being only partially coupled at low temperature, but rather the influence of thermal noise. At low temperatures, the GB displacements are very small, on the same order of magnitude as the thermal noise. During computation, this noise can be misinterpreted as actual GB displacements, leading to inaccuracies in the calculated $\rho^2$ and shear coupling strengths. This issue, detailed in Section S3 of the Supplementary Materials, can be mitigated by extending the simulation time for each individual interface random walk simulation.

## 4. Discussion

### 4.1 The advantages and applications of the 3D-IRW theory and FAIRWalk method

*4.1.1 Elimination of the external driving force effect*

Most investigations [30,34,77,78] into the shear coupling behavior of GBs have relied on the application of external driving forces and measured the ratio of velocities or displacements in different directions. However, as demonstrated in a previous study [30], GB shear coupling is highly sensitive to the external driving force. In Chen et al.'s work [30], the apparent shear coupling factor β approaches zero or infinity as the temperature increases, depending on the type of driving force applied, as shown in Fig. 8. Similarly, Yu et al. [77] observed this trend using the ramped synthetic driving force (RSDF) method. The study [30] indicates that the apparent GB shear coupling behavior depends on the type and magnitude of the applied driving force, leading to the conclusion that shear coupling is not an intrinsic GB property. However, the extremely high driving forces used in atomistic simulations can drive the GB far from equilibrium [44], resulting in unrealistic outcomes.

Shear coupling strengths determined from 3D-IRW simulations eliminate the influence of external driving forces. As shown in Fig. 8, compared with results obtained under external driving forces [30,77], $S_{zx}$ closely aligns with the value predicted by disconnection theory[30,57] (denoted by the gray dotted line) within the fully coupled temperature range (800K-1300K). Additionally, previous research [53] has shown that treating GB shear coupling as a tensor, rather than a single value like the shear coupling factor $\beta$, allows Eqs. 2 and 28 to accurately predict variations in apparent shear coupling under different external forces. Therefore, while the apparent shear coupling, i.e. $\beta$, depends on the driving force, the shear coupling strengths and the associated shear coupling tensor should be regarded as intrinsic properties of the GB.

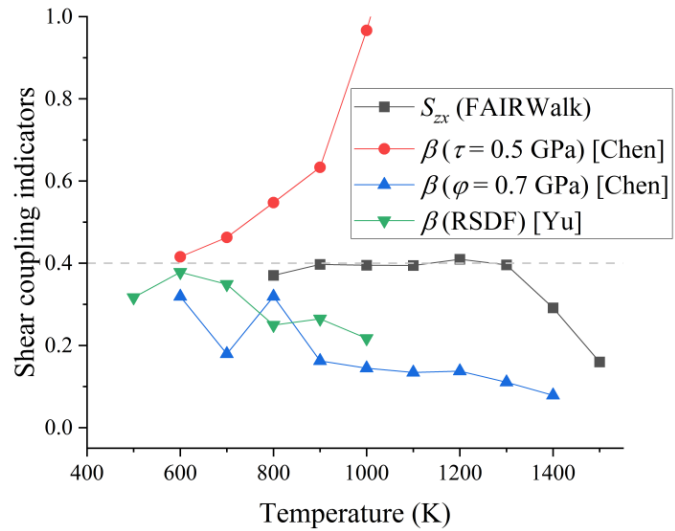

Figure 8 Comparison of various shear coupling indicators for the Σ13 (5 1 0) GB, determined using different methods. Shear coupling strengths ($S_{zx}$) are obtained via the FAIRWalk method, while $\beta$ vs. $T$ curves are replotted from the studies by Chen et al. [30] and Yu et al. [77]. The dashed line indicates the shear coupling of the disconnection mode with minimum activation energy.

*4.1.2 Reflection of the intrinsic GB shear coupling state*

As shown in Fig. 9a, for the Σ17 (2 0 0) GB (p17 in Olmsted database [58]), the mobilities in the *y* and *z* directions vary synchronously with temperature. If one evaluates the shear coupling state in these two directions using the ratio of velocities ($v_y/v_z$) or displacements ($Y/Z$), they may calculate a specific shear coupling factor. However, the near-zero standard coupling factor $\rho^2$, which represents the product of the shear coupling strengths, i.e. $S_{yz} S_{zy}$, indicates no coupling between the two directions. This lack of coupling is also reflected in the plot of GB displacements in the *y* and *z* directions (inset in Fig. 9a), where the displacements are completely uncoupled (for comparison, Fig. 5a shows an example of fully coupled GB displacements).

Fig. 9b shows that the dichromatic pattern of the Σ17 (2 0 0) GB is symmetric in the *y* and *z* directions, which explains the synchronous mobilities. However, the disconnection modes either lack shear coupling or involve two modes with opposite shear coupling directions. As a result, GB migrations in the *y* and *z* directions are independent, as reflected by the $\rho^2$ value. This feature, aside from the Σ3 (2 2 2) GB (p3 in the Olmsted database [58], which remains immobile across all temperature ranges), is observed in all {1 1 1} and {1 0 0} twist GBs in the database, accounting for 7.2% of the 388 CSL GBs.

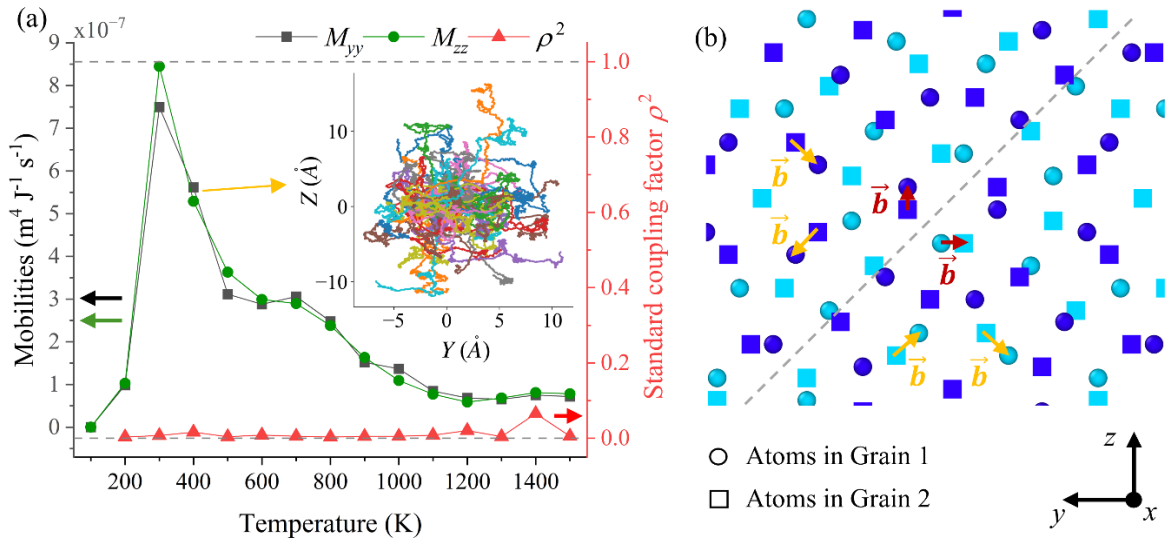

Figure 9 (a) Variation of $M_{yy}$, $M_{zz}$ and $\rho^2$ with temperature for the Σ17 (2 0 0) GB. The inset shows the plot of GB displacements in the *y-z* plane at 400K. (b) Dichromatic pattern analysis for the Σ17 (2 0 0) GB, with arrows indicating the Burgers vectors of potential disconnection modes.

## 4.2 Survey of 388 Ni CSL GBs

Building on the advancements of 3D-IRW theory and the FAIRWalk method, we conducted an extensive survey of GB mobility tensors and the associated shear coupling for 388 Ni CSL GBs from the Olmsted database [58]. All mobility tensors were determined using Eq. 26, while shear coupling strengths were derived from Eq. 27 (or Eq. 28). To minimize the influence of thermal noise, we excluded shear coupling strengths associated with very small mobilities ($M < 10^{-9}$ m$^4$/(J·s)) and set $\rho$ to zero for $\rho^2 < 0.06$. See Section S3 in the Supplementary Materials for further details. Additionally, activation energies for GB migrations were calculated by fitting GB mobilities to the classical thermal activation model [22,45], with the computational method [44] detailed in Section S4 in the Supplementary Materials. The computed GB mobility tensors and corresponding activation energies are provided in Appendix A and illustrated in Appendix B. Shear coupling strengths and the standard coupling factor $\rho^2$ are listed in Appendix C and illustrated in Appendix D.

The following section highlights some intriguing phenomena uncovered during the survey.

### *4.2.1 Temperature dependent GB shear coupling behavior*

According to the disconnection theory [57], an increase in temperature activates more disconnection modes, leading to a gradual weakening of GB shear coupling until it eventually disappears, as shown in Fig. 7. This trend has been confirmed by numerous previous studies [24,25,30] and is consistent with most GBs in our survey. However, we observed an opposite trend in the Σ7 (5 4 1)/(5 4 $\bar{1}$) GB (p26 in the Olmsted database [58]). As shown in Fig. 10a, this GB exhibits no shear coupling below 1100K, but a sudden appearance of shear coupling in the *x-z* plane occurs at 1200K. This emergence coincides with a sharp decline in GB normal mobility ($M_{xx}$) and the beginning of an increase in GB shear mobility in the *z* direction ($M_{zz}$). A similar phenomenon has also been observed in Σ55 (12 3 1)/(12 3 $\bar{1}$) and Σ37 (11 10 1)/(11 10 $\bar{1}$) GBs (p219 and p382 in the Olmsted database [58]) Neither GB roughening transition [44,57] nor topological transition [79] adequately explains this phenomenon. A potential explanation could be a structural phase transition [80–86], however, to verify this, more advanced methods for determining GB phases at high temperatures (1000K and above) are necessary. Another possible explanation involves atomic disruptive jumps in the GB area. Previous research [87] has shown that disruptive jumps in the coordinated movement of GB atoms can cause stagnation of the GB.

At high temperatures, the excess volume between GB atoms increases, leading to a higher frequency of disruptive jumps. These cumulative jumps may force the GB to alter its migration pattern, explaining the sudden drop in normal mobility.

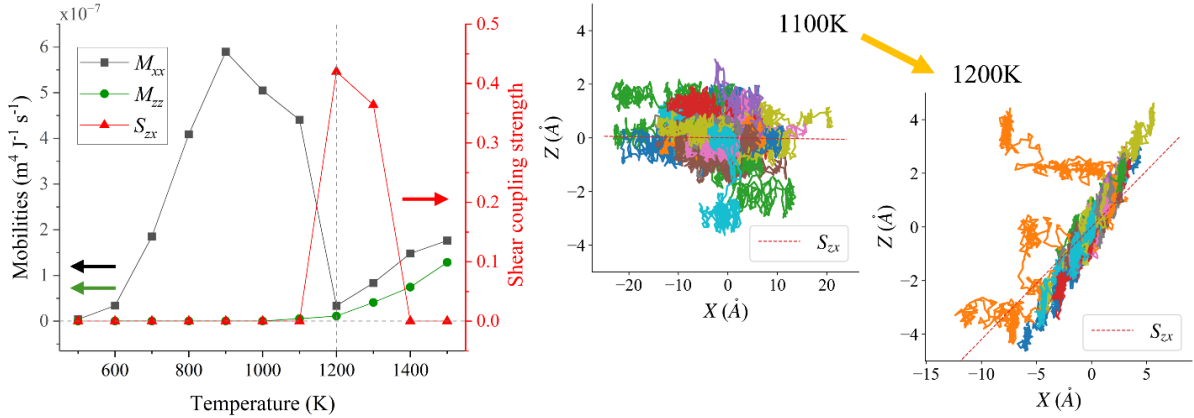

a. Temperature-induced emergence of shear coupling (Σ7 (5 4 1)/(5 4 $\bar{1}$))

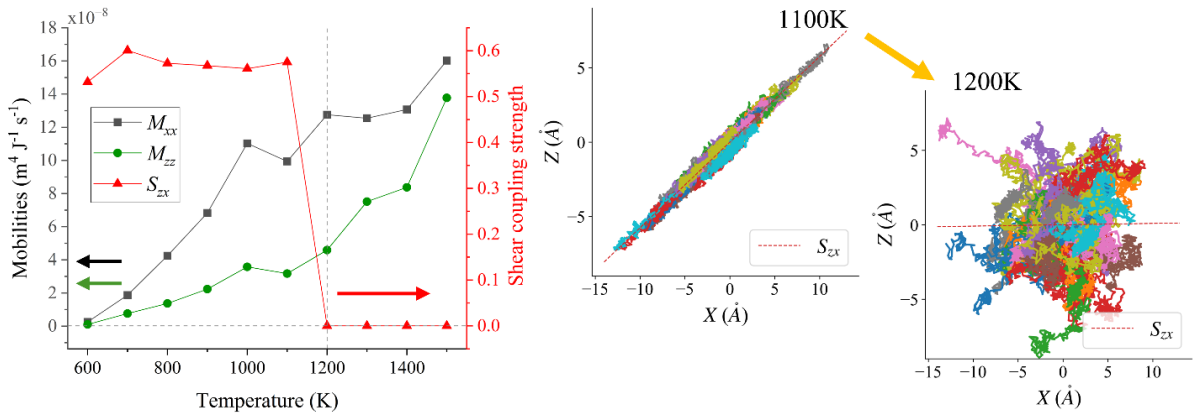

b. Temperature-induced disappearance of shear coupling (Σ39 (4 3 1)/(4 3 $\bar{1}$))

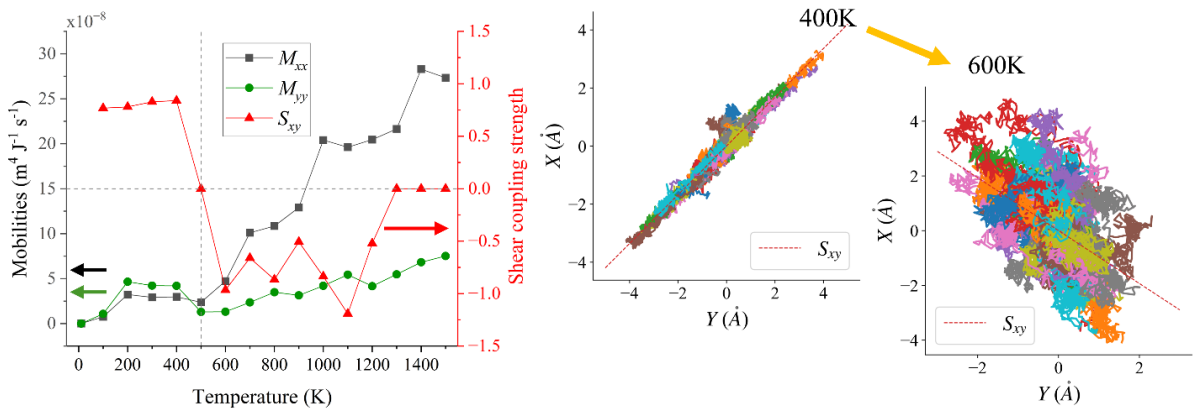

c. Temperature-induced change in shear coupling direction (Σ45 (11 2 1)/(11 2 $\bar{1}$))

Figure 10 Temperature-induced shear coupling phenomena in Σ7 (5 4 1)/(5 4 $\bar{1}$), Σ39 (4 3 1)/(4 3 $\bar{1}$), and Σ45 (11 2 1)/(11 2 $\bar{1}$) GBs.

In contrast, we observed a sudden disappearance of shear coupling in the Σ39 (4 3 1)/(4 3 $\bar{1}$) GB (p70 in the Olmsted database [58]). As shown in Fig. 10b, the GB transitions abruptly from fully coupled to completely uncoupled migration between 1100K and 1200K. Given the rapid nature of this transition, rather than the gradual disappearance of shear coupling shown in Fig. 7, a topological transition [79] is a possible explanation. In this process, the long-range interaction between disconnections vanishes at a certain temperature, allowing for the sudden activation of multiple disconnection modes. This type of transition is usually accompanied by an increase in GB mobility, which aligns with the trend in Fig. 10b.

We also observed a temperature-induced change in the direction of shear coupling. As illustrated in Fig. 10c, when the temperature rises from 400K to 600K, the Σ45 (11 2 1)/(11 2 $\bar{1}$) GB (p192 in the Olmsted database [58]) undergoes a transition from fully coupled, to uncoupled, and then to partially coupled in the opposite direction. Unlike previous studies [56], where constraints were applied, this change in shear coupling direction appears to be purely thermally driven.

These phenomena challenge the current understanding of GB shear coupling and warrant further investigation.

*4.2.2 "Immobile" GBs*

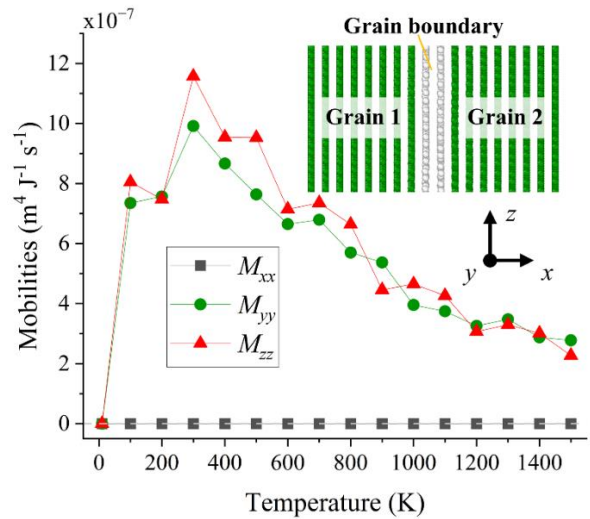

Figure 11 Variation of GB normal and shear mobilities with temperature for Σ37 (2 2 2) GB. The inset shows the layered configuration of GB atoms in the normal (*x*) direction.

Some GBs in the survey exhibit near-zero or even "zero" normal mobility ($M_{xx}$) across all temperatures. An example is the Σ37 (2 2 2) GB (p388 in the Olmsted database [58]), shown in

Fig. 11. It is important to note that, in the interface random walk method, even minimal local atomic vibrations can produce variations in GB position and corresponding mobility, as discussed in Section S2 of the Supplementary Materials. Therefore, "zero" mobility indicates that even normal atomic vibrations are undetectable. These types of GBs make up 7% of the Olmsted database [58]. A common feature among them is that they are {111} and {100} twist GBs, with their CSL plane lying in the *y-z* plane (the {111} or {100} GB plane), and they exhibit a layered configuration of GB atoms in the direction normal to GB plane, as shown in the inset of Fig. 11. This layered structure restricts normal GB atomic jumps while promoting GB sliding, which explains the low normal mobility but relatively high shear mobility. Moreover, Fig. 11 shows that the shear mobilities exhibit typical non-Arrhenius (or anti-thermal) behavior, suggesting relatively low energy barriers for shear motion in this GB [44,45]. Therefore, although traditionally these GBs would be classified as "immobile"[34,43], in response to external stresses or material deformation, GB sliding and grain rotation are more likely to occur in them.

*4.2.3 Activation energies for GB migration*

Figure 12a illustrates the distribution of activation energies for GB normal migration across 388 GBs, while the activation energies associated with shear mobilities are provided in Fig. S9 of the Supplementary Materials. As shown in Fig. 12a, most Σ3 GBs exhibit exceptionally low activation energies. According to the classical thermal-activation model [22,45], the transition temperature at which GB mobility shifts from a thermally activated trend (where GB mobility increases with temperature) to an anti-thermal trend (where GB mobility decreases with temperature) is proportional to the activation energy [44]. This implies that non-Arrhenius (or anti-thermal) behavior is more likely to occur in GBs with low activation energy for migration. As shown in Fig. 12a, most Σ3 GBs and a noticeable proportion of Σ7 and Σ9 GBs exhibit extremely low activation energies. This explains why the anti-thermal trend in GB migration is predominantly observed in Σ3[33,49,88,89], Σ7[43,90,91], and Σ9[43] GBs.

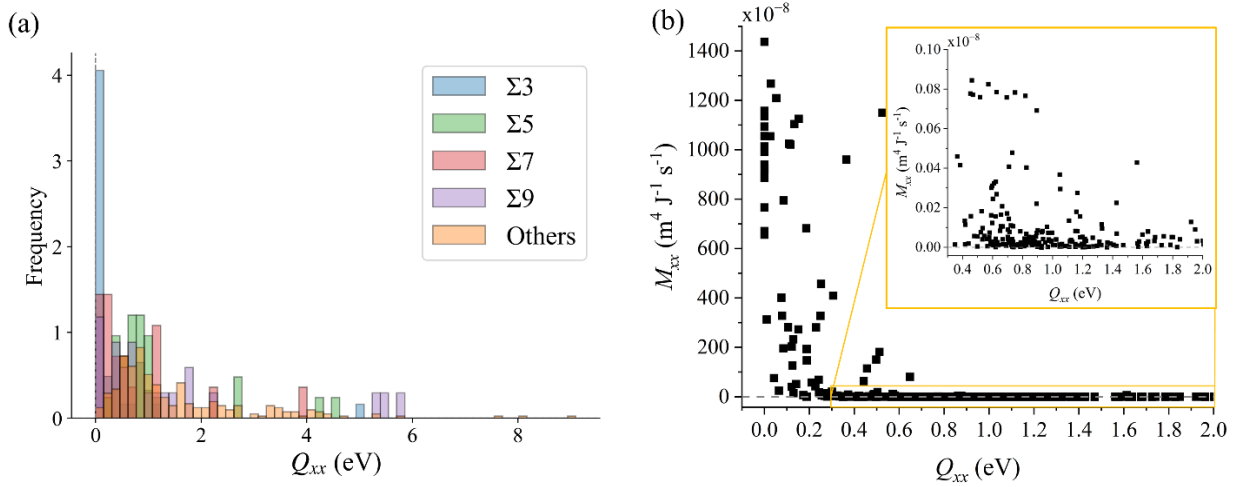

Figure 12 (a) Distribution of activation energies for normal GB migration in Σ3, Σ5, Σ7, Σ9, and other GBs in the Olmsted database [58]. (b) Plot of GB normal mobility vs. activation energy.

Figure 12b shows the relationship between activation energy (0 to 2 eV) and GB normal mobility at room temperature (300K). While it generally aligns with the expectation that high mobility GBs are concentrated in the low activation energy range, no clear linear relationship between the two variables can be established. According to disconnection theory [46,57], additional factors, such as the height ($h$) and Burgers vector ($\vec{b}$) of the disconnection modes—which determine the distance of each hop during GB migration—also play a significant role in influencing GB mobility. Verifying this would require more detailed information on the disconnection modes for each GB.

**4.3 Limitations of the equilibrium methods**

A critical condition for applying equilibrium methods to extract GB mobility is that the GBs must exhibit movement or fluctuations at the target temperatures. Previous studies [52,53,92], along with the FAIRWalk method we proposed, have focused on enhancing the efficiency of these methods—specifically, extracting more accurate GB mobility from less data. However, if a GB does not move (e.g., at low temperatures), equilibrium methods cannot yield useful information. This is the primary limitation of these methods, which is why they are mainly applied at high temperatures [40,93] or for GBs with low activation energy for migration [44].

When tracking the variance of GB displacement (e.g., $\sigma_x^2$) over time, a long relaxation period is often observed before the $\sigma_x^2$-$t$ relationship stabilizes. This relaxation period is more pronounced in GBs with a large energy barrier for migration. For instance, the symmetric Σ5 (3 1 0) GB, known for large activation energy, shows no significant random walk behavior until 1100K. However, as illustrated in Fig. 13, even at a lower temperature of 800K, a glass-like relaxation period is evident, characterized by ballistic, caging, and diffusive stages. A similar phenomenon was reported by Zhang et al. [94], who observed these stages in the mean square displacement of individual atoms in the GB region, identifying them as characteristic of glassy material dynamics. These findings suggest that the overall GB dynamics exhibit similar behavior. Soltani et al. [95] proposed that this relaxation time is linked to GB migration dynamics. This feature holds promising potential to overcome the limitations of traditional equilibrium methods, paving the way for extracting GB mobility at low temperatures, even before random walk behavior initiates.

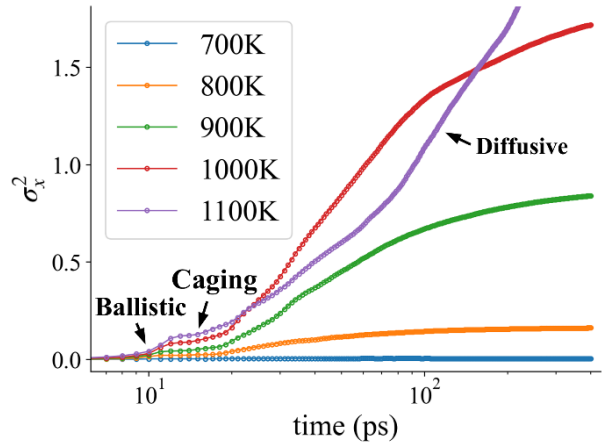

Figure 13 Plot of the variance of GB displacement ($\sigma_x^2$) over time for Σ5 (3 1 0) GB from 700K to 1100K. The GB exhibits negligible movement from 700K to 1000K, with evident random walk behavior emerging at 1100K.

## 5. Conclusion

The main conclusions of the current study are as follows:

1. 3D-IRW theory. We extended the conventional interface random walk theory from 1D to 3D. The key results are encapsulated in Eq. 9 or 22, which enable the extraction of the 3D GB mobility and shear coupling tensors from interface random walk simulations. The computed GB mobility and shear coupling tensors capture the most intrinsic properties of the GB, which are unaffected by the excessively large driving forces commonly used in atomistic simulations. Additionally, we mathematically proved the symmetry of the GB mobility tensor in the case of overdamped GB migration, a characteristic confirmed by MD simulation results.

2. FAIRWalk method. Building on the 3D-IRW theory, we refined the FAIRWalk method [53] to efficiently extract the GB mobility tensor with fewer simulations, while maintaining high accuracy.

3. Survey of 388 Ni CSL GBs in the Olmsted Database. A vast survey was conducted on the GB mobilities, shear coupling, and activation energies of 388 GBs. Several intriguing phenomena were uncovered, including temperature-induced sudden emergence, disappearance, or inversion of shear coupling; GBs with "zero" normal mobility but high shear mobility; and a non-linear relationship between activation energy and mobility. These findings challenge conventional understanding of GB migration and warrant further investigation. The complete survey results are presented in Appendices A to D.

**Acknowledgment**

The authors thank Dr. David L Olmsted for sharing the 388 Ni GB structure database. This research was supported by NSERC Discovery Grant (RGPIN-2019-05834), Canada, and the use of computing resources provided by Research Alliance of Canada. X.S. also acknowledges financial support from the University of Manitoba Graduate Fellowship (UMGF). During the preparation of this manuscript the authors used ChatGPT to improve its readability. After using this tool, the authors reviewed and edited the manuscript as needed and take full responsibility for the content of the publication.

Supplementary materials

for

# Intrinsic grain boundary mobility tensor from three-dimensional interface random walk

Xinyuan Song and Chuang Deng*

Department of Mechanical Engineering, University of Manitoba, Winnipeg, MB R3T 2N2, Canada

* Corresponding author: Chuang.Deng@umanitoba.ca

**S1 The choice of timestep for molecular dynamics (MD) simulations**

In the Verlet algorithm [1], the position vector $x(t)$ at the moment $t + \Delta t$ is given by

$$x(t + \Delta t) = 2x(t) - x(t - \Delta t) + a(t)\Delta t^2 + \mathcal{O}(\Delta t^4) \tag{S1}$$

where, $\Delta t$ is the timestep, $a(t)$ is the acceleration, and $\mathcal{O}(\Delta t^4)$ is truncation error which is proportional to $\Delta t^4$. Generally, a timestep of 5 fs ($5 \times 10^{-15}$ s) is much smaller than the atomic vibration period in a nickel crystal (approximately $10^{-13}$ s), making truncation errors negligible in most cases. However, when the system dynamics is too subtle, such as in low-temperature scenarios, the truncation errors can become pronounced. For instance, on a 2D potential energy surface, as illustrated in Fig. S1a, when the system dynamics are near the center of a potential energy well, and the system starts from one side of the well with zero initial velocity, the restoring force pulls it toward the center. However, if the timestep is too large, the system may overshoot to the opposite side of the well, and potentially reach a higher energy state [2].

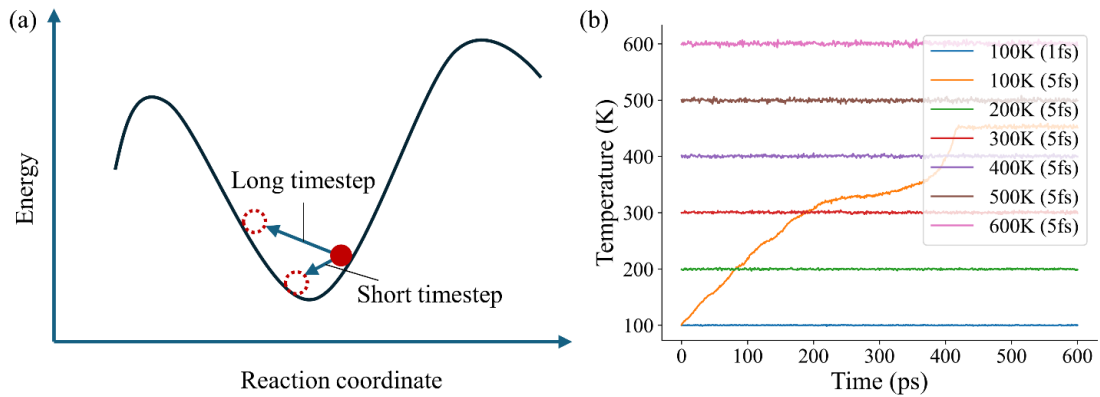

Figure S1 (a) Illustration of truncation errors caused by a large timestep. (b) Temperature variation during interface random walk simulations in the NVE ensemble.

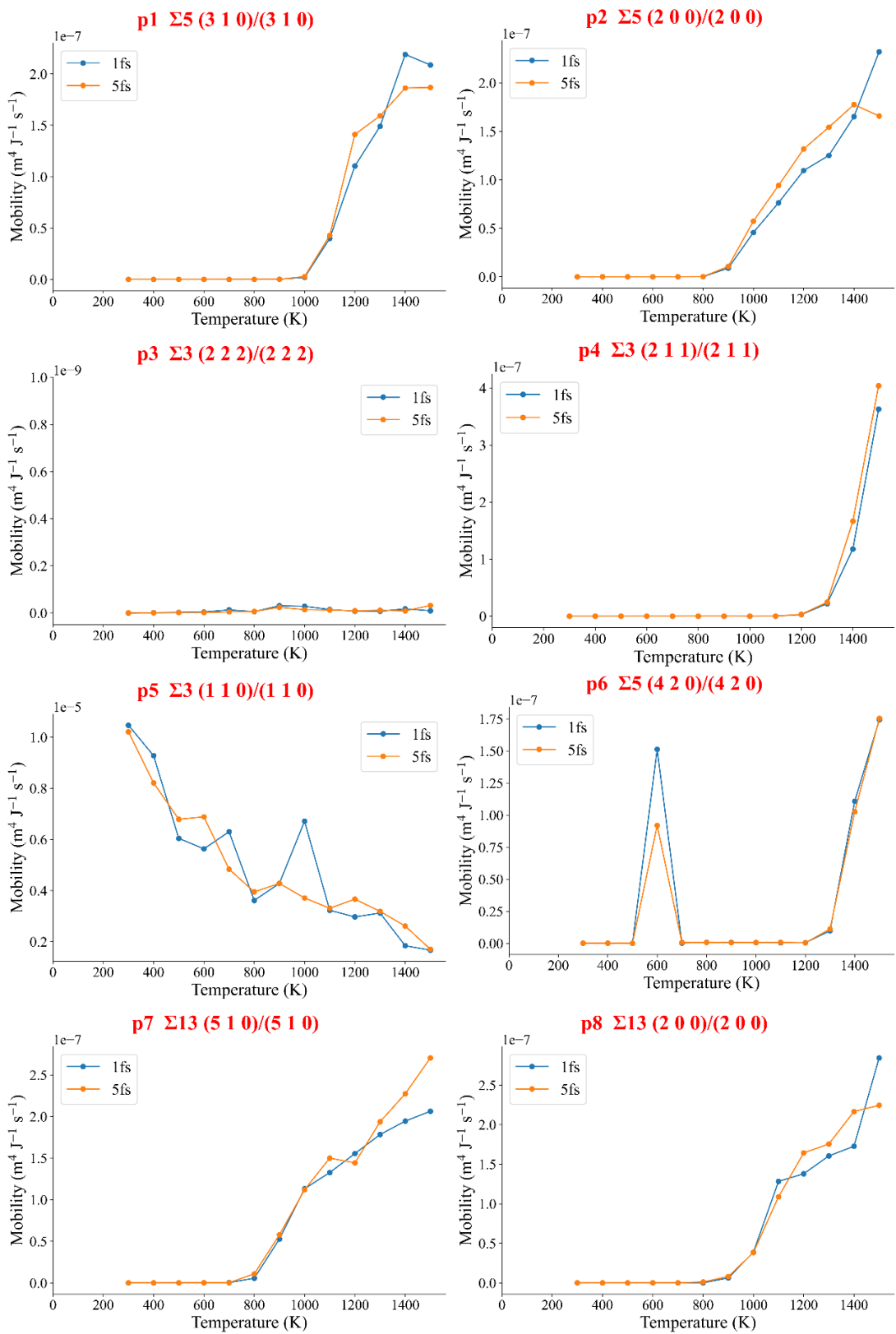

Figure S2: Comparison of the computed GB normal mobility ($M_{xx}$) from 300K to 1500K for p1 to p8 in the Olmsted database [3], using different timesteps.

Ref. [2] suggests that simulations at lower temperatures, where dynamics are slower, matches smaller timesteps. As shown in Fig. S1b, during interface random walk simulations under the microcanonical (NVE) ensemble, a timestep of 5 fs at 100K causes an increase in system temperature, violating energy conservation. However, this issue does not occur at 200K and higher, and can be resolved by using a smaller timestep of 1 fs.

To balance accuracy and computational efficiency, we apply two different timesteps for simulations in distinct temperature ranges: 1 fs for the low-temperature range (10K-200K) and 5 fs for the room-to-high temperature range (300K-1500K). Fig. S2 compares the computed grain boundary (GB) normal mobility from 300K to 1500K for p1 to p8 in the Olmsted database [3] using different timesteps. The differences fall within normal fluctuations.

## S2 Computation of GB mobility tensor

The GB mobility tensors for 388 GBs in Olmsted database [3] are computed as follows:

$$\begin{bmatrix} Var(X) & Cov(X,Y) & Cov(X,Z) \\ Cov(X,Y) & Var(Y) & Cov(Y,Z) \\ Cov(X,Z) & Cov(Y,Z) & Var(Z) \end{bmatrix} = \frac{k_B T t}{A} \begin{bmatrix} M_{xx} & M_{xy} & M_{xz} \\ M_{yx} & M_{yy} & M_{yz} \\ M_{zx} & M_{zy} & M_{zz} \end{bmatrix}^T \qquad (S2)$$

where $X$, $Y$, and $Z$ are cumulated displacements of the GB in the $x$, $y$, and $z$ directions, respectively. The computed mobility tensors for 388 Ni GBs are provided in Appendix A, with an illustration of the results in Appendix B.

It is important to note that in the low-temperature regime, the calculated mobilities for some GBs are extremely low, based on the variation of their average position under thermal fluctuations. These fluctuations are primarily due to the local vibrations of GB atoms, as shown in Fig. S3, and do not indicate actual GB random movement. Therefore, in this study, GB mobilities lower than $10^{-9}$ m$^4$/(J s) are considered immobile and are excluded from further analysis related to shear coupling and activation energy.

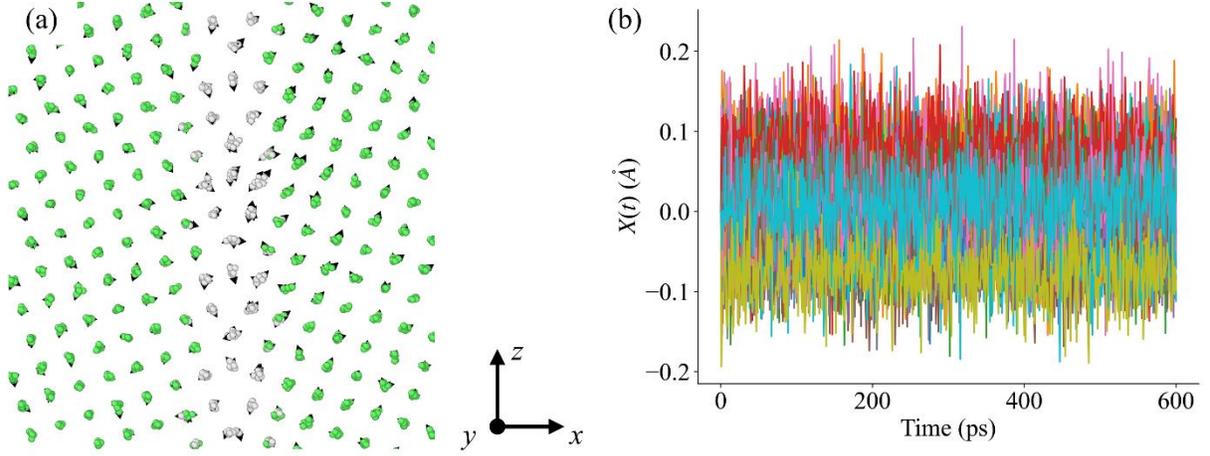

Figure S3 (a) Illustration of local atomic vibrations in the Σ5 (3 1 0) GB (p1 in Olmsted database [3]) at 500K, and (b) the resulting variance in the GB position.

**S3 Computation of shear coupling strengths and shear coupling tensor**

The shear coupling strengths are determined by linearly fitting the cumulative GB displacements from interface random walk simulations. The relationship between the shear coupling strengths (or the GB shear coupling tensor) and the GB mobility tensor is described in Eq. S3:

$$\begin{bmatrix} M_{xx} & M_{xy} & M_{xz} \\ M_{yx} & M_{yy} & M_{yz} \\ M_{zx} & M_{zy} & M_{zz} \end{bmatrix} = \begin{bmatrix} 1 & S_{xy} & S_{xz} \\ S_{yx} & 1 & S_{yz} \\ S_{zx} & S_{zy} & 1 \end{bmatrix} \begin{bmatrix} M_{xx} & & \\ & M_{yy} & \\ & & M_{zz} \end{bmatrix} \quad (S3)$$

In this extensive study, we calculated all six shear coupling strengths from Eq. S3. To reveal clear variation trends, two filters were applied:

1) Exclude results associated with immobile boundaries ($M < 10^{-9}$ m$^4$/(J s)). The shear coupling strength is expressed as:

$$S_{\alpha\beta} = \frac{M_{\alpha\beta}}{M_{\beta\beta}} = \rho \frac{\sigma_\alpha}{\sigma_\beta} \quad (S4)$$

As shown in Fig. S4a, the Σ35 (4 2 0) GB (p57 in Olmsted database [3]) is nearly immobile in the *y*-direction but demonstrates noticeable mobility in the *x*- and *z*-directions. Although there is no coupling in the *x-y* or *z-y* planes ($\rho$ is small), a large ratio between the variation of GB displacements in the *x* or *z* directions to that in the *y*-direction can still result in significant values for $S_{xy}$ or $S_{zy}$. This could skew the analysis of the survey results on GB shear coupling. To eliminate such misleading results, the data are filtered using the following criterion: if $M_{\beta\beta}$ in Eq. S4 is less than $10^{-9}$ m$^4$/(J s) (considered immobile), then $S_{\alpha\beta}$ is marked as "not available" or "-" in Appendix C. As shown in Fig. S4c, after applying this filter, only results reflecting true shear coupling trends remain.

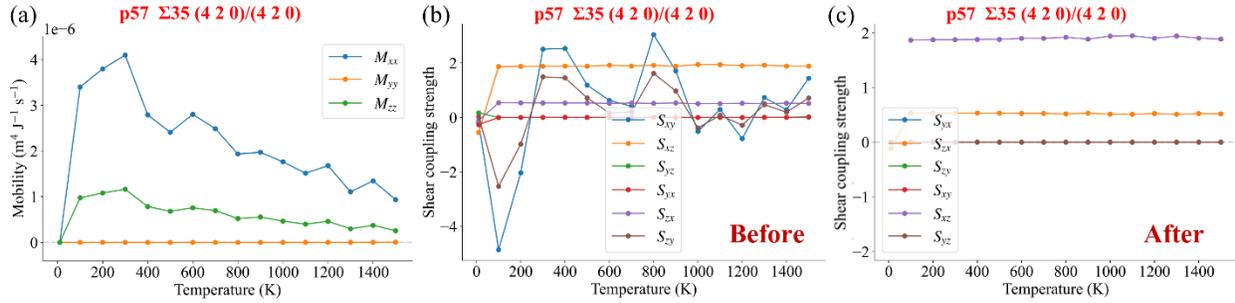

Figure S4 (a) Variation of principal mobilities with temperature for the Σ35 (4 2 0) GB. (b, c) Variation of computed shear coupling strengths with temperature before and after filtering the results associated with immobile mobilities.

2) Zeroing out small $\rho$ values: Some GBs exhibit a very weak coupling effect, as indicated by a $\rho^2$ value close to zero, as shown in Fig. S5a. However, due to the extremely large scale factor $\sigma_\alpha/\sigma_\beta$, the computed shear coupling strengths show significant fluctuations, which can skew the analysis, as illustrated in Fig. S5b. To mitigate this issue, all $\rho$ values with $\rho^2<0.06$ were set to zero.

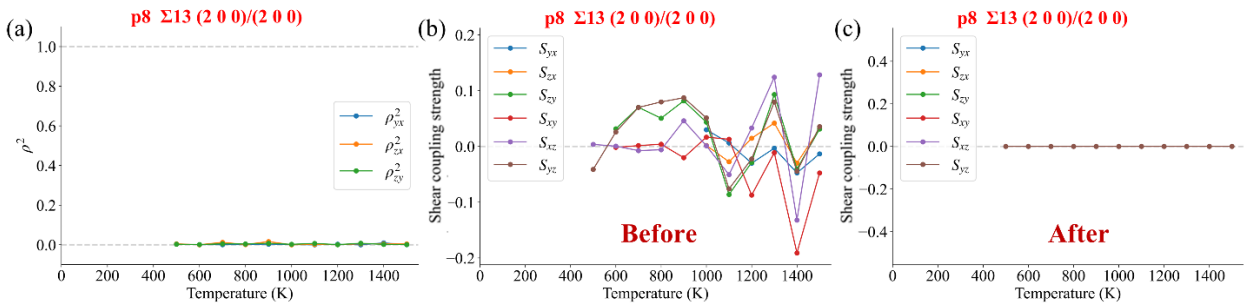

Figure S5 (a) Variation of coupling factor $\rho^2$ with temperature for Σ13 (2 0 0) GB (p8 in Olmsted database [3]). (b, c) the variation of computed shear coupling strengths with temperature before and after zeroing out small $\rho$ values.

After applying the two filters, most GBs exhibit clear *S-T* trends, revealing many interesting phenomena. In addition to the shear coupling strengths, we also calculated the coupling factor $\rho^2$, a normalized value ranging from 0 to 1 that indicates the coupling state of the GBs. The detailed results are provided in Appendix C and illustrated in Appendix D.

It is worth noting the phenomenon of an intriguing 'initial rise' in $\rho^2$ at low temperatures, as shown in Fig. S6a. This phenomenon is commonly observed throughout Appendix D and is caused by thermal noise at low temperatures when the GB is small. This thermal noise can be misinterpreted as GB displacements, leading to inaccuracies in the computed coupling factor ($\rho^2$) and shear coupling strengths, as shown in Fig. S6b. As the temperature increases, the thermal noise becomes negligible compared to actual GB displacements, allowing the computed $\rho^2$ and shear coupling strengths to more accurately reflect the true shear coupling state, as shown in Fig. S6c.

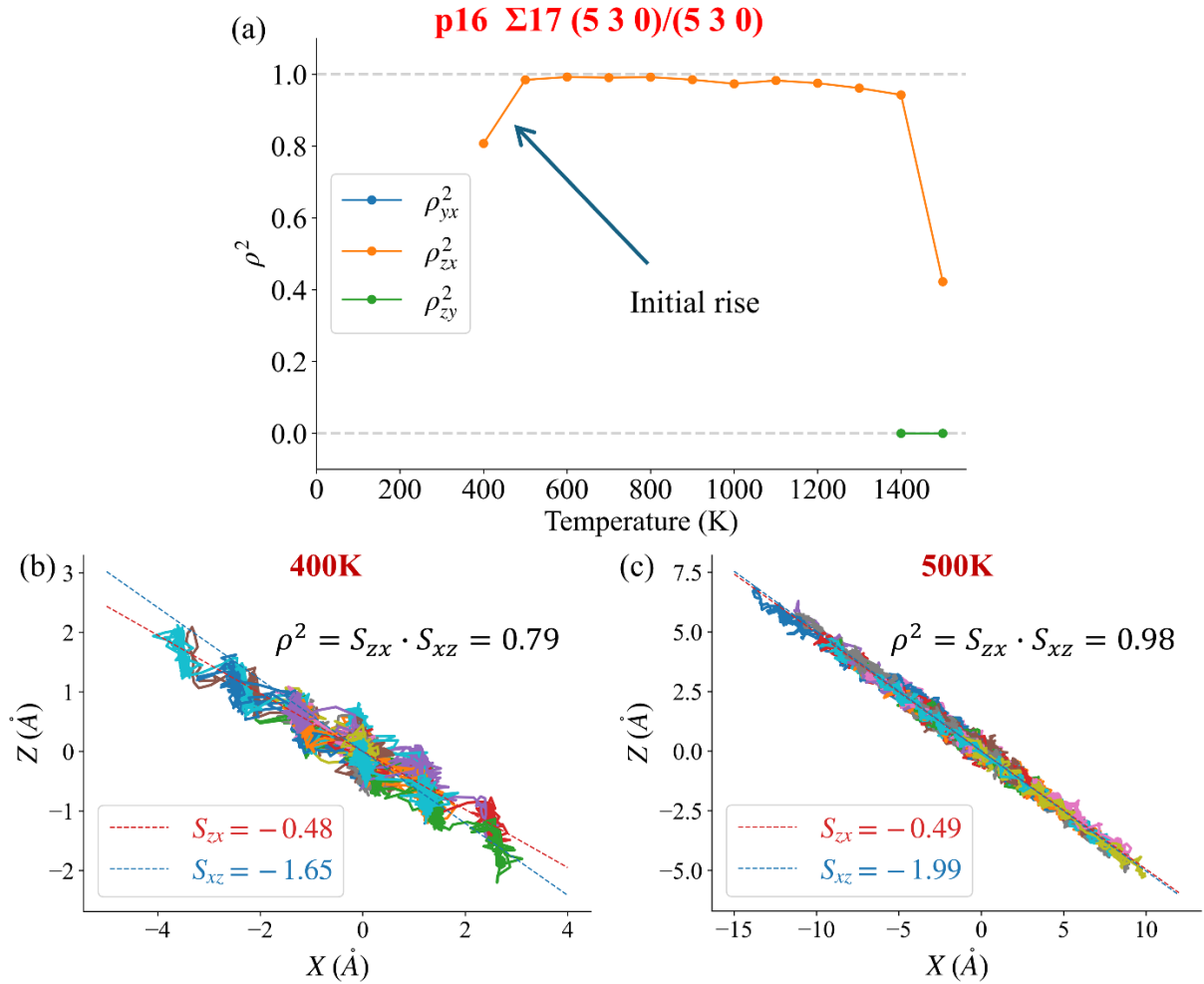

Figure S6 (a) The "initial rise" in $\rho^2$ at low temperatures. (b-c) Illustration of the effect of thermal noise on computed shear coupling strengths and coupling factor $\rho^2$.

## S4 Calculation of activation energies

According to the classical thermal-activation model [4,5], the GB migration velocity is determined by the frequency ($\omega$) of GB jumping forward (+) and backward (-) on the potential energy surface:

$$v = Nb \left[ \omega^+ \exp\left(-\frac{Q}{k_B T}\right) - \omega^- \exp\left(-\frac{Q+\Psi}{k_B T}\right) \right]$$

$$= Nb\omega \exp\left(-\frac{Q}{k_B T}\right)\left[1 - \exp\left(-\frac{\Psi}{k_B T}\right)\right] \quad (S5)$$

where, $Q$ is activation energy for GB migration, $\Psi$ is external driving force, $k_B$ is Boltzmann constant, $N$ is the number of atoms, and $b$ is the distance of the atom jump. In the zero driving force limit, Eq. S5 simplifies to [4,6]

$$M = \frac{v}{P} = \frac{Nb\omega C}{k_B T} \exp\left(-\frac{Q}{k_B T}\right) \quad (S6)$$

Eq. S6 describes the non-Arrhenius behavior of GB mobility: $M$ initially increases with temperature and then decreases after reaching a peak at the transition temperature $T_{trans}$, as shown in Fig. S3a. By setting the derivative of Eq. S6 with respect to $T$ to zero, i.e., $M'=0$, we obtain [6]

$$Q = k_B T_{trans} \quad (S7)$$

Eq. S7 demonstrates a linear relationship of $Q$ and $T_{trans}$. However, this equation assumes that GB migration is mediated by a single disconnection mode, which is valid for most GBs only in the low-temperature regime. At higher temperatures, the activation of multiple modes and factors such as phase transitions and roughening transitions can influence the apparent thermal behavior of GB mobility, leading to inaccuracies in determining $T_{trans}$.

By taking the natural logarithm of both sides of Eq. S6, one can derive [6]:

$$\ln M = -\frac{Q}{k_B T}\left(1 + \frac{k_B T \ln T}{Q}\right) + C \quad (S8)$$

where $C$ is a constant. When $k_B T \ln T \ll Q$, i.e. $T \ll T_{trans}$, the terms in the bracket becomes 1, and the GB mobility exhibits Arrhenius behavior, as shown by the red curve in Fig. S7a.

In our survey of the mobility tensors of 388 Ni GBs, most of GBs exhibit the $\ln M$-$1/(k_B T)$ curves similar to the trend in Fig. S7b: at low temperatures, the GB remains stationary, but $\ln M$ gradually increases with temperature due to the growing intensity of local atomic vibrations in the GB region. As the temperature increases further, the GB begins to move, and $\ln M$ and $1/(k_B T)$ display a linear relationship over a certain temperature range. At high temperatures, the $\ln M$-$1/(k_B T)$ curve deviates from Arrhenius behavior.

In our study, we determine the activation energies $Q$ for GB migration by calculating the maximum slope of the $\ln M$-$1/(k_B T)$ curves, as illustrated in Fig. S7b. The results are listed in Appendix A.

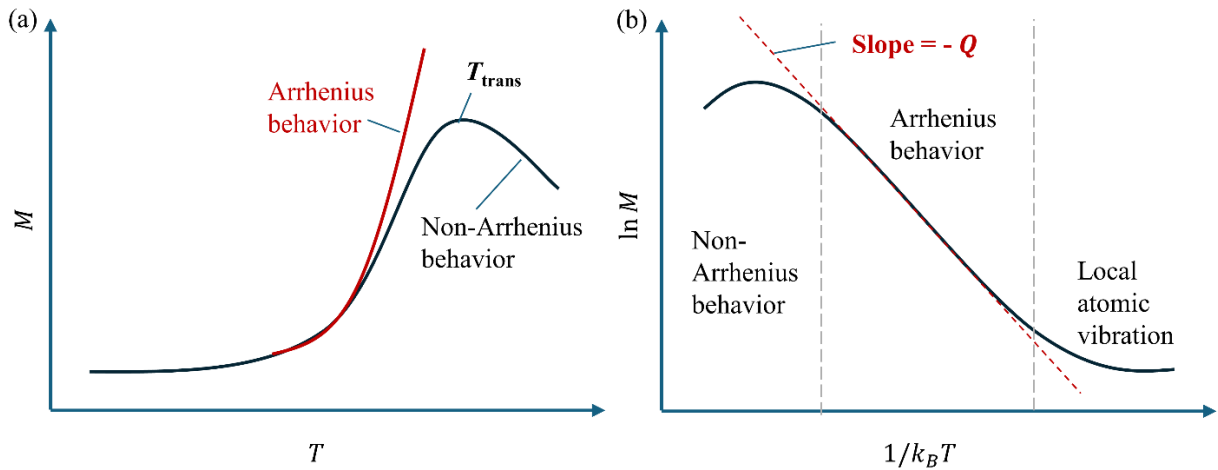

Figure S7 Illustration of (a) the $M$-$T$ curve as described by Eq. S6 and (b) the $\ln M$-$1/(k_B T)$ curve as described by Eq. S8.

It is important to note that some GBs with extremely low mobility do not move at any temperature in the random walk simulations. In such cases, the fitted $Q$ value would be meaningless. Therefore, we only calculated activation energies for GBs with mobilities greater than $10^{-9}$ m$^4$/(J s). Additionally, some GBs, such as Σ111 (11 10 1) (p381 in Olmsted database [3]), exhibit an anti-thermal trend across the entire tested temperature range, as shown in Fig. S8. According to Eq. S7, these GBs likely have extremely low activation energies for GB migration, so we approximate $Q$ for these GBs as zero.

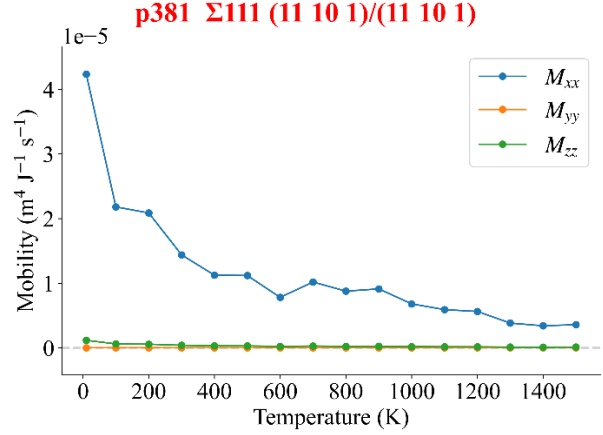

Figure S8 Illustration of the anti-thermal mobility trend across the entire tested temperature range.

Olmsted's 388 Ni GB database [3] includes 41 Σ3 GBs (10.57%), 27 Σ5 GBs (6.96%), 19 Σ7 GBs (4.9%), and 23 Σ9 GBs (5.93%). These four types of CSL GBs have been extensively studied in previous research [7–9]. We plotted the distribution of $Q$ across different Σ values for these GBs. The distribution of activation energy associated with GB normal mobilities (i.e., $M_{xx}$) is shown in Fig. 12a of the main article, while the distributions of activation energies associated with shear ($M_{yy}$ and $M_{zz}$) and coupled mobilities (i.e., $M_{yx}$, $M_{zx}$, and $M_{zy}$) are illustrated in Fig. S9.

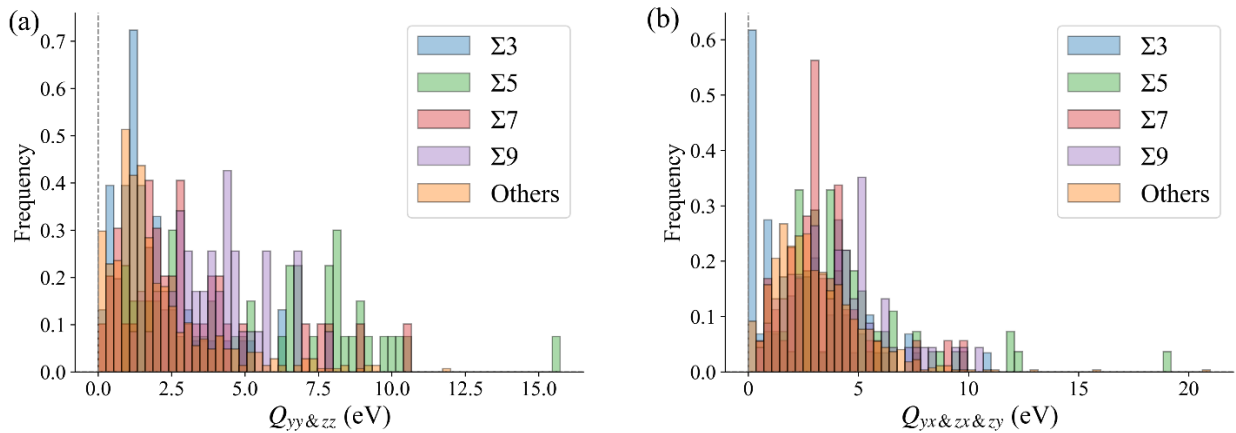

Figure S9 Distributions of activation energies associated with shear and coupled mobilities.